# Comparative Study of Long Short-Term Memory (LSTM) and Quantum Long Short-Term Memory (QLSTM): Prediction of Stock Market Movement

Tariq Mahmood[1*], Ibtasam Ahmad[1], Malik Muhammad Zeeshan Ansar[1], Jumanah Ahmed Darwish [2], Rehan Ahmad Khan Sherwani [3]

[1] Centre for High Energy Physics, University of the Punjab, Lahore, Pakistan. Email: tariqmahmood.chep@pu.edu.pk
[2] Department of Statistics, Faculty of Science, University of Jeddah, Jeddah, Saudi Arabia. Email: jadarwish@uj.edu.sa
[3] College of Statistical Sciences, University of the Punjab, Lahore-Pakistan. Email: rehan.stat@pu.edu.pk
*Corresponding Author

**Abstract**

In recent years, financial analysts have been trying to develop models to predict the movement of a stock price index. The task becomes challenging in vague economic, social, and political situations like in Pakistan. In this study, we employed efficient models of machine learning such as long short-term memory (LSTM) and quantum long short-term memory (QLSTM) to predict the Karachi Stock Exchange (KSE) 100 index by taking monthly data of twenty-six economic, social, political, and administrative indicators from February 2004 to December 2020. The comparative results of LSTM and QLSTM predicted values of the KSE 100 index with the actual values suggested QLSTM a potential technique to predict stock market trends.

## 1. INTRODUCTION

For developing nations, sustaining macroeconomic stability is a primary challenge [1]. The stock market's pivotal role in redistributing financial resources among diverse economic entities is widely recognized. Consequently, progress within the stock market echoes advancements in a country's economic growth trajectory [2, 3]. This interrelation is evident as the stock market's movements reflect a nation's economic health—positive stock market performance signifies growth, while negative trends signal otherwise. Hence, it is important to identify the economic factors that affect stock market variations since they impact the country's stock market movement. Previous studies indicate that the stock market capitalization rate, influenced by currency rates, gross domestic product, current account, interest rates, and money supply, has a major impact [4, 5].

Hashmi and Chang examined E7 stock indices to show the effect of macroeconomic variables across different states of stock markets—bullish, bearish, and normal [6]. The study's outcomes revealed noteworthy trends. In the long term, trade balance, foreign direct investment, and industrial production index emerged as significant influencers of emerging stock indices. Moreover, employing the QARDL model, the research demonstrated that the short-term effects of factors such as consumer price index, foreign direct investment, interest rate, and exchange rate exhibit variations in different market states. Notably, the long-term effect shows variability for all macroeconomic variables except for the industrial production index [6].

Moreover, stock prices interlinked to the equilibrium position in the long run by 18.6% adjustment speed through the channel of FDI, GDP, and provision of domestic loans to the private sector. The study's results also showed that while financial development has a negatively influenced stock prices in the short term, FDI has a long-term favorable impact [7]. Another noteworthy economic indicator is the exchange rate. Consequently, Chang et al. revisited the complex associations among oil prices, real exchange rates, and stock market prices within China [6, 8]. Their findings illuminate varying connections between oil and stock prices and between stock prices and exchange rates, dependent on diverse quantile combinations [6]. However, exchange rate fluctuations in China, India, and the USA illustrate the minimal influence on the daily closing price of stock indices such as SSE, Nifty50, and DJI, respectively. These variations considerably impact the number of shares traded on each of the three stock exchanges, as revealed by Krishnan and [9].

Asymmetry exists in changes to the money supply, industrial production, and real exchange rate (RER) on stock returns, and the asymmetries are more pronounced after the 2002 subsample than for the entire sample period. The empirical findings imply that easy monetary policy enhances stock returns more than restrictive monetary policy [10]. Similarly, understanding how foreign reserves relate to the stock market is crucial because, recently, building up international reserves has been the preferred strategy taken by emerging nations to ensure financial stability [11]. According to [12], the stock market is positively impacted by economic sentiment indices and market capitalization. In comparison, exports and industrial production have encouraging relationships with stock prices. Inflation has demonstrated a negative link with stock prices [13].

In addition to macroeconomic indicators, the administrative quality of a nation plays a pivotal part in shaping its performance in the stock market [14, 15]. Aligned with regulatory and legal institutions and Finance-Law theories, they substantially influence the stock market by shaping financial and economic activities [16]. In their attempt to comprehend how corruption control and government integrity shake the long-term growth of Pakistan's stock market, Islam et al. discovered that these factors favorably influence the pace of the country's equity market [17]. Their investigation indicates that government integrity and effective corruption control contribute positively to the progression of Pakistan's equity market. Notably, aspects such as instances of bankruptcy, contractual misconduct, inflexible or stringent securities laws, inadequacies of the legal system, levels of expropriation, uncertainty in property safety laws, and the inadequacy of prosecution agencies all manifest as methods of exploitation, collectively impeding the seamless operation of the equity market [18].

The pragmatic findings provide the theoretical claim that indeterminate socio-political circumstances harm economic evolution in the Greek paradigm and show a robust damaging relationship between indeterminate socio-political states and the overall index of the Athens Stock Exchange (ASE). Since political unpredictability has a detrimental impact on market values, an unstable political system will eventually cause stock prices to fall [13].

In the work conducted by Asongu, a robust and noteworthy positive connection is defined between measures of performance of the stock market and the eminence of government institutions [19]. To consider the dynamics of capitalization, value traded, turnover, and the number of listed companies on the stock market, researchers looked at the dynamics of government effectiveness in terms of corruption control, government efficiency, political stability or lack of violence, voice and accountability, regulation quality, and rule of law. These dynamics were instrumented using data on income levels, levels of press freedom, religious dominance, and legislative precedents. The research revealed that countries with more advanced governmental institutions would favor stock markets with bigger market capitalization, higher share prices, better turnover ratios, and more listed enterprises [19].

A larger market capitalization is used to calculate the value of stock sets. A few technical elements may be employed to obtain statistical data from the value of stock prices [20]. Stock indexes frequently evaluate each country's economic status based on the prices of stocks with significant market investment [21]. Because of the uncertain nature of stock price fluctuation, the movement of stock values is ambiguous. Additionally, governments typically have trouble determining the status of the market. Since stock values are typically nonlinear, non-parametric, and dynamic, they frequently result in poor statistical model performance and make it difficult to predict precise values and movements [22].

The projection of stock group prices has long been interesting and challenging for investors due to its non-linearity, inherent dynamism, and complexity. Both economists and computer scientists are interested in stock market forecasting since it is a traditional but challenging subject. To develop an effective prediction model, linear and machine-learning approaches have been studied over the past 20 years. Deep learning techniques have recently been suggested as new directions for this subject and issue. Forecasting stock market trends is a significant work that needs close attention since, with the appropriate judgments, a good prediction of stock prices may lead to the possibility of acquiring attractive rewards. Because of the difficult problem of stock market forecasting and the data's chaotic, noisy, and non-stationary nature, the forecast causes investors to pause and consider investing in future benefits [23]. By using monthly data from 26 economic, social, political, and administrative indicators from February 2004 to December 2020, we were able to predict the Karachi Stock Exchange (KSE) 100 index using effective machine learning models like long short-term memory (LSTM) and quantum long short-term memory (QLSTM). The KSE 100 index's anticipated values from LSTM and QLSTM and their actual values imply that QLSTM may be able to forecast stock market developments.

Furthermore, various empirical studies have scrutinized the influence of macroeconomic variables on stock prices [24, 25]. However, the existing body of literature presents inconsistent findings, primarily focusing on developed countries while lending less attention to their developing counterparts. Hence, a directed spotlight on Pakistan becomes imperative, particularly employing contemporary techniques like QLSTM to prognosticate the KSE100 index. This approach is warranted due to the divergent risk and return dynamics characterizing developing economies in contrast to their developed counterparts [26, 27].

## 1.1. Long Short-Term Memory (LSTM)

Sepp Hochreiter and Jürgen Schmidhuber introduced Long Short-Term Memory (LSTM) in 1997 [28, 48]. It is a form of recurrent neural network (RNN) and can model complex sequential data by preserving long-term memory and selectively apprising information. It is intended to overcome the vanishing gradients that arise in conventional RNNs [29-33]. Deep learning has become a fundamental building block, particularly in tasks involving a sequence of data, such as natural language processing, time series analysis, and speech recognition [33-39]. In sequential data, LSTMs can capture and remember long-range dependencies to effectively model sequences with complex dependencies and patterns. For the storage and retrieval of information, LSTM networks consist of different memory cells [28]. Three gates—the input, forget, and output gates—control the information flow across the LSTM network. LSTMs can effectively handle both short-term and long-term dependencies in the data [49]. These gates selectively forget and update the information from the past [28]. It may involve the following steps:

1. Data reparation:- splitting data into training, validation, and testing sets.
2. Data encoding:- conversion of input data into a suitable format for the LSTM.
3. Model architecture (specification of the number of layers, neurons in each layer, activation function, etc.).
4. Model compilation (selection of loss function (e.g., mean squared error for regression or categorical cross-entropy for classification), selection of an optimizer (e.g., Adam, RMSprop) to minimize the loss function, monitoring of additional metrics like accuracy or mean absolute error)
5. Training (feed training dataset to LSTM network, compute loss function and gradients, update model weights with the help of optimizer and backpropagation,

repeat the process until the model converges)
6. Validation (monitor its generalization ability on the validation dataset)
7. Testing (asses its real-world performance by feeding unseen data)
8. Post-processing (conversion of predicted probabilities into class labels for classification tasks)
9. Deployment (make real-time predictions by integrating the trained LSTM model into the application/system)
10. Fine-Tuning and Maintenance (fine-tune your LSTM model with new data to ensure it remains accurate and up to date).

**1.2. Quantum Long Short-Term Memory (QLSTM)**

Quantum Long Short-Term Memory (Q-LSTM) builds on LSTM's legacy. In modern deep learning, LSTM is a cornerstone for natural language processing and time series analysis [29-39][47]. Q-LSTM, a quantum-inspired variant, leverages quantum bits (qubits) for data storage and processing. It harnesses quantum principles, like superposition and entanglement, for enhanced efficiency. Q-LSTM models long-range dependencies, surpassing classical LSTMs [40-45]. Quantum gates control information flow. In summary, Q-LSTM offers quantum-level precision, ideal for complex sequential data tasks, marking a significant advancement in deep learning. QLSTMs proficiently manage both short-term and long-term data dependencies [44]. Quantum gates within QLSTM selectively adapt information from the past, akin to traditional LSTMs [40].

The QLSTM workflow entails:
1. Data Preparation: Divide data into training, validation, and test sets to facilitate model training and evaluation.
2. Data Encoding: Transform input data into a suitable format for QLSTM, considering quantum encoding methods.
3. Model Architecture: Give a description of the design, including the number of layers, the number of neurons in each layer, the activation mechanisms, and any quantum-inspired improvements.
4. Model Compilation: Specify the loss function (e.g., mean squared error or quantum-specific variants), choose an optimizer (e.g., quantum-inspired optimizers), and monitor additional metrics such as quantum fidelity.
5. Training: Feed the training dataset to the QLSTM, compute loss, update weights through quantum backpropagation, and iterate until convergence, harnessing quantum parallelism.
6. Validation: Consider quantum validation techniques to assess the model's generalization ability on the validation dataset.
7. Testing: Evaluate QLSTM's real-world performance using unseen data, employing quantum testing strategies.
8. Post-processing: For classification tasks, convert quantum-aided predicted probabilities into class labels.
9. Deployment: Seamlessly integrate the trained QLSTM model into applications/systems for real-time quantum-enhanced predictions.
10. Fine-Tuning and Maintenance: Continuously fine-tune the QLSTM model with new quantum data to ensure its accuracy and relevance in evolving quantum computing landscapes.

**2. Material and Methods**

This study has employed monthly data spanning from February 2004 to December 2020. The analysis involved twenty-six distinct, independent variables to predict Pakistan's stock market dynamics. These variables encompassed a wide spectrum, including balance of trade, consumer financing for house building, consumer price index (representing inflation), control of corruption, crude oil prices, domestic savings, exchange rate, external debt stocks, foreign direct investment, foreign exchange reserves, GDP growth rate, gold price, government effectiveness, households final consumption expenditure, industrial production index, industry value added, labor force participation rate, money supply, personal remittances growth, political stability and absence of violence/terrorism, portfolio investment, growth rate, regulatory quality, the rule of law, three-month treasury bill rates, and wholesale price index. The dependent variable, in this case, was the closing price of the KSE 100 index, serving as a measure of the stock market's performance. The data for these variables were drawn from reputable sources, such as the World Development Indicator, the State Bank of Pakistan, and the Pakistan Bureau of Statistics. An artificial neural network was employed to interpolate data, converting monthly data into daily observations.

We have utilized LSTM and QLSTM to predict the KSE 100 index based on the above indicators and compared their results.

## 2.1. Long Short-Term Memory (LSTM) Algorithm

The Long Short-Term Memory (LSTM) algorithm involves matrix operations and activations. The block diagram of LSTM is shown in the following Figure 1.

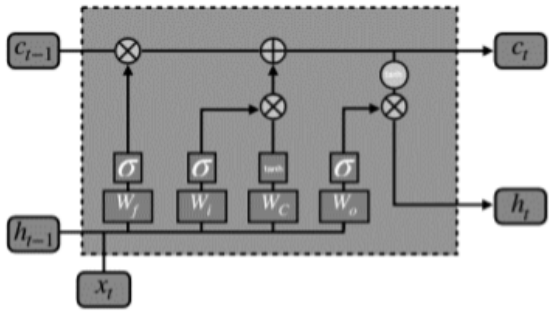

Figure 1: Block diagram of long-short-term memory (LSTM) describing the information flow through different gates with different activation functions and corresponding weight matrices [46].

A high-level description of the LSTM algorithm in terms of mathematical operation and data flow is as:

1. Initialize the LSTM cell state (C) and hidden state (h) with zeros or small random values.
2. Calculate the input gate activation ($i_t$) by applying a sigmoid function to a weighted sum of the current input ($x_t$) and the previous hidden state ($h_{t-1}$). This gate determines which information from the current input and previous hidden state should be stored in the cell state.
$$i_t = \sigma([h_{t-1}, x_t] * w_i + b_i)$$
3. Apply a sigmoid function to the weighted sum of the current input ($x_t$) and the prior hidden state ($h_{t-1}$) to determine the forget gate activation ($f_t$). The information from the cell state ($C_{t-1}$) that should be forgotten or maintained is decided by this gate.
$$f_t = \sigma([h_{t-1}, x_t] * w_f + b_f)$$
4. Apply the hyperbolic tangent (tanh) activation function to a weighted sum of the current input ($x_t$) and the prior hidden state ($h_{t-1}$) to get the candidate's cell state ($C_{tilde}$). New potential values for the cell state are computed in this stage.
$$C_{tilde} = tanh([h_{t-1}, x_t - x_{t-1}] * w_c + b_c)$$
5. The candidate cell state ($C_{tilde}$), the prior cell state ($C_{t-1}$)), and the forget gate ($f_t$) are combined to update the cell state ($C_t$)). In this stage, the data that should remain in the cell state is decided.
$$C_t = i_t * C_{tilde} + f_t * C_{t-1}$$
6. Apply a sigmoid function to the weighted sum of the current input ($x_t$) and the prior hidden state ($h_{t-1}$) to determine the output gate activation ($O_t$). The information from the cell state that should be produced as the hidden state ($h_t$) is decided by this gate.
$$O_t = \sigma([h_{t-1}, x_t] + w_o + b_o)$$
7. Apply the hyperbolic tangent activation function (tanh) to the updated cell state ($C_t$)

multiplied by the output gate ($O_t$) to get the new hidden state ($h_t$). Information that will be sent to the following time step and may be used as the final output is included in this concealed state.

$$h_t = O_t * tanh\,(C_t)$$

8. The hidden state ($h_t$) can be used as the output of the LSTM cell for the current time step, or it can be passed to subsequent layers in a deep LSTM network.
9. Repeat the above steps for each time step in the sequence.

Where:
- $i_t$ is the input gate activation.
- $\sigma$ is the sigmoid activation function.
- $w_i$ is the weight matrix for the input gate.
- $x_t$ is the current input.
- $h_{t-1}$ is the previous hidden state.
- $b_i$ is the bias of the input gate.
- $f_t$ is the forget gate activation.
- $w_f$ is the weight matrix for the forget gate.
- $b_f$ is the bias for the forget gate.
- $C_{tilde}$ is the candidate cell state.
- $b_c$ is the bias for the candidate cell state.
- tanh is the hyperbolic tangent activation function.
- $w_c$ is the weight matrix for the candidate cell state.
- $C_t$ is the updated cell stat.
- $C_{t-1}$ is the previous cell state.
- $o_t$ is the output gate activation.
- $w_o$ is the weight matrix for the output gate.
- $b_o$ is the bias of the output gate.
- $h_t$ is the new hidden state.

The LSTM algorithm captures and propagates information over long sequences by allowing the network to update and forget information selectively.

## 2.2. Quantum Long Short-Term Memory (QLSTM) Algorithm

Certainly, Quantum Long Short-Term Memory (Q-LSTM) builds upon LSTM's legacy by infusing quantum computing principles into its architecture. While LSTM remains a stalwart in deep learning for sequential data tasks, Q-LSTM represents a quantum leap forward, offering enhanced computational efficiency and the capacity to masterfully capture both short-term and long-term dependencies within data, all within the realm of quantum-inspired computing. The block diagram of QLSTM is shown in the following Figure 2.

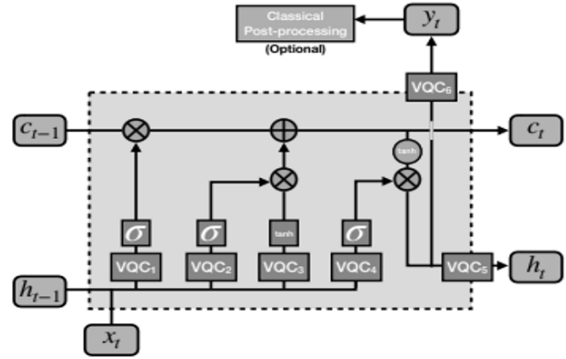

Figure 2: Block diagram of quantum long-short-term memory (QLSTM) describing the information flow through different gates with different activation functions along with variational quantum circuits (VQCs) [46].

Here's a brief overview of each line in the simplified Quantum Long Short-Term Memory (Q-LSTM) algorithm code:

1. Import necessary libraries:
   - Imports essential libraries, such as NumPy for numerical operations and Qiskit for quantum programming.
2. Define quantum circuit parameters:
   - Sets the number of qubits in the quantum circuit (`n_qubits`) and the number of classical bits used for measurement (`n_bits`).
3. Initialize the quantum circuit:
   - Creates a quantum circuit using Qiskit, specifying the number of qubits and classical bits.
4. Define Q-LSTM quantum gates:
   - Defining the quantum bit and classical bit in the co-relation for the maximum probability 1. This streamlined approach helps pave the way for the subsequent layers of our process.
- Defines quantum gates for the input gate, forget gate, candidate cell state, and output gate, each with specific quantum operations.

$$i_t = \sigma(x_t W_{xi} + h_{t-1} W_{hi} + b_i)$$
$$f_t = \sigma(x_t W_{xf} + h_{t-1} W_{hf} + b_f)$$
$$C_t = tanh(x_t W_{xc} + h_{t-1} W_{hc} + b_c)$$
$$O_t = \sigma(x_t W_{xo} + h_{t-1} W_{ho} + b_0)$$

Where:
- $i_t$ is the input gate at time step t
- σ is the sigmoid activation function
- $W_{xi}$ and $W_{hi}$ are weight matrices
- $x_t$ is the input vector at time step t
- $h_{t-1}$ is the hidden state at time step t−1
- $b_i$ is a bias vector
- $f_t$ is the forget gate at time step t
- $W_{xf}$ and $W_{hf}$ are weight matrices
- $b_f$ is a bias vector
- $C_t$ is the candidate cell state at time step t
- tanh is the hyperbolic tangent activation function
- $W_{xc}$ and $W_{hc}$ are weight matrices
- $b_c$ is a bias vector
- $O_t$ is the output gate at time step t
- $W_{xo}$ and $W_{ho}$ are weight matrices
- $b_0$ is a bias vector

5. Quantum LSTM operation for one time step:
    - Calls the functions defined earlier for the input gate, forget gate, candidate cell state, and output gate, combining these gates to represent one Q-LSTM time step (VQC).

$$i_t = \sigma(VQC_2(v_t))$$
$$f_t = \sigma(VQC_1(v_t))$$
$$\tilde{C}_t = tanh(VQC_3(v_t))$$
$$o_t = \sigma(VQC_4(v_t))$$
$$C_t = f_t * C_{t-1} + i_t * \tilde{C}_t$$
$$y_t = VQC_6(tanh\, tanh(C_t) * o_t)$$

Where:
- $v_t$ is the N-dimensional input vector $v_t = (x1, x2, \cdots, xN)$
- VQC1 and VQC2 control the flow of information between the input and output gates.
- VQC3 and VQC4 control the flow of information between the input and forget gates.
- VQC5 and VQC6 control the flow of information between the input and update gates.

Variational quantum circuits (VQCs) are a type of hybrid quantum-classical algorithm. VQCs work by taking advantage of quantum mechanics' superposition and entanglement properties. Superposition allows VQCs to represent multiple possible solutions to a problem simultaneously. Entanglement allows VQCs to share information between different parts of the circuit. Here, VQCs are checking the probability of occurring. VQC circuits vary from problem to problem; therefore, one of the circuits is shown in the following figure.

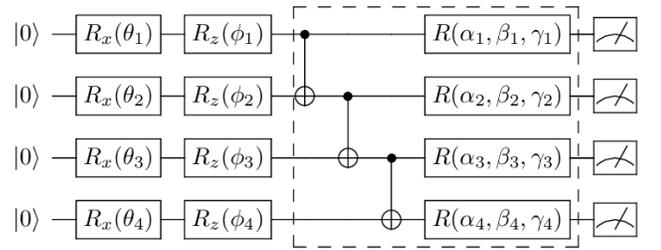

Figure 3: Generic variational quantum circuit architecture for the deep Q network [46]

6. Example usage of quantum LSTM:
    - Generates random quantum input ($x_t$) and random previous quantum hidden state ($h_{prev}$) for a single time step. In practice, these values would be real data.

$$x_t = \sum_{i=1}^{n} \alpha_i |x_i\rangle$$

$$h_{prev} = \sum_{i=1}^{n} \beta_i |h_i\rangle$$

Where:
- $x_t$ represents a state or vector at time tt.
- n is the number of terms in the summation.
- $\alpha_i$ represents the coefficients or weights associated with each $|x_i\rangle$ term.
- $|x_i\rangle$ represents a quantum state or vector

associated with the i-th term.
- $h_{prev}$ represents a state or vector called the "previous hidden state."
- $\beta_i$ represents the coefficients or weights associated with each $|h_i\rangle$ term.
  $|h_i\rangle$ represents a quantum state or vector associated with the i-th term.
7. Define weight matrices and biases for gates:
   - Defines quantum circuits for weight matrices ($w_i$, $w_f$, $w_c$, $w_o$) and biases ($b_i$, $b_f$, $b_c$, $b_o$) for the input, forget, candidate cell state, and output gates.
8. Perform a quantum LSTM step:
   - Calls the `quantum_lstm_step` function to apply one Q-LSTM time step to the quantum circuit, incorporating the quantum gates, input, and biases.
9. Measure quantum circuit to obtain output:
   - Adds measurement operations to the quantum circuit to obtain classical measurement results from the qubits.
10. Simulate the quantum circuit:
    - Select a quantum simulator (Qiskit's Aer simulator) to execute the quantum circuit and simulate quantum measurements.
11. Execute the quantum circuit and obtain results:
    - Executes the quantum circuit and obtains measurement results (`counts`) from the simulator.
12. Display measurement results:
    - Prints the measurement results obtained from the quantum circuit, representing the outcome of the Q-LSTM step.

This code provides a simplified illustration of a Q-LSTM step using a quantum framework. Developing a complete Q-LSTM model would require more complex implementations, optimized quantum gates, and integration with quantum hardware or advanced simulators for practical applications.

### 3. Results and Discussion

One of the most lucrative issues in modern finance is the accurate forecast of future stock values, which may result in significant profit and reduced risk. Recurrent neural networks (RNNs) with Long Short-Term Memory (LSTM) may be applied to sequential data processing and classification problems. Because of this, many individuals have had great success using LSTM to predict future stock values using sequences of previous data.

On the other hand, recent research has demonstrated that the LSTM may be made more effective and trainable by swapping out part of its layers for variational quantum layers. Therefore, we obtain a hybrid quantum-classical LSTM model, abbreviated QLSTM for quantum LSTM. A study [44, 50] shows that QLSTM is more trainable than its classical counterpart because it learns local features more efficiently and significantly more data after the first training epoch while utilizing similar parameters. Considering these most recent findings, we continue to test our variational quantum-classical hybrid neural network approach on stock price projections.

In the following notebook, we provide a proof of concept for applying QLSTM in stock price prediction, demonstrating that it can produce results that are on par with or even superior to those of its traditional counterpart. To do this, we use the same number of features in both LSTM and QLSTM to predict the stock prices of the KSE 100 index.

We have gathered historical information on the KSE stock prices, emphasizing the closing price. Using LSTM (or QLSTM), we aim to predict KSE's closing stock prices.

We have gathered the following relevant data and information to accomplish this goal. We have chosen not to go into detail about the data collected in this study, even though it is fascinating and significant in and of itself. We divided the data into 80% for training purposes and 20% for training to validate the accuracy.

Both models undergo a training phase, during which they learn the underlying patterns by adjusting their internal parameters. This involves feeding the historical data into the models, comparing their predictions against the actual output, and optimizing their parameters to minimize the prediction error. Figure 4 and Figure 5 show the mean squared error loss of both LSTM and QLSTM during training and testing, respectively.

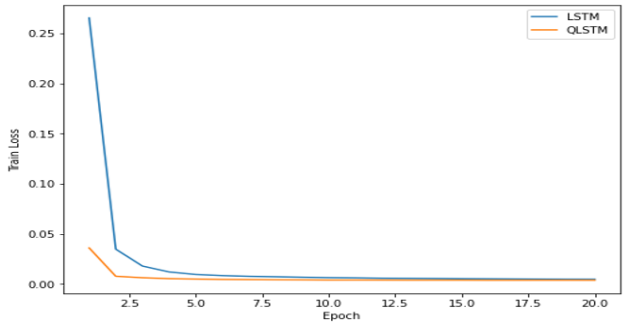

Figure 4: *Mean squared error loss during training of LSTM and QLSTM networks.*

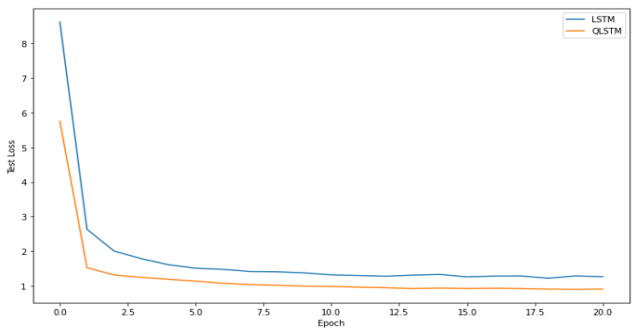

Figure 5: Mean squared error loss during testing of LSTM and QLSTM networks.

Once trained, both models are ready for predictions. Given a set of new input features, they can project how these features relate to the output column based on the patterns they've learned during training. Figure 6 shows the LSTM's prediction of the close value of the KSE 100 index, and Figure 7 shows the QLSTM's prediction of the close value of the KSE 100 index.

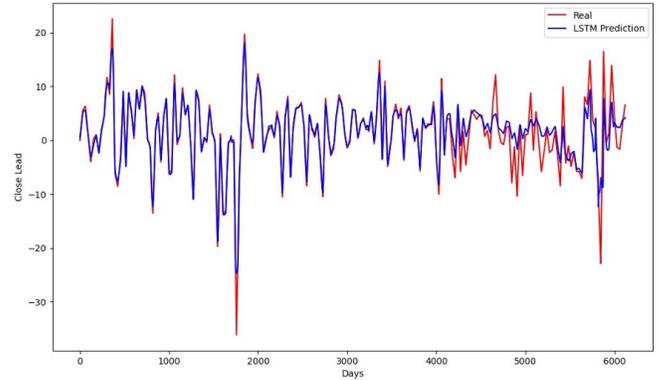

Figure 6: Comparison of real data of the KSE 100 index and the prediction of LSTM.

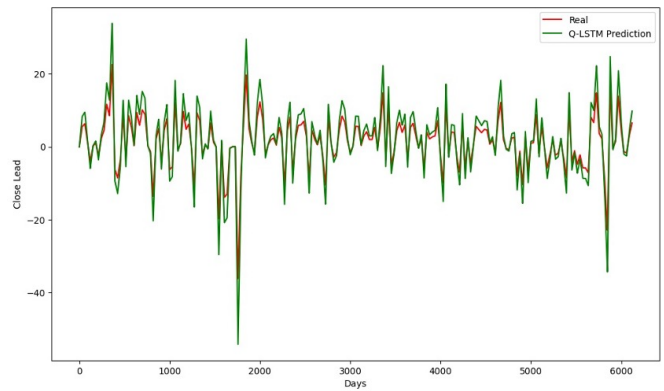

Figure 7: Comparison of real data of KSE 100 index and the prediction of QLSTM.

Figure 8 compares the LSTM's predictions and QLSTM's predictions of the KSE 100 index for the initial data of 1000 days. Figure 9 compares the LSTM's and QLSTM's predictions of the KSE 100 index.

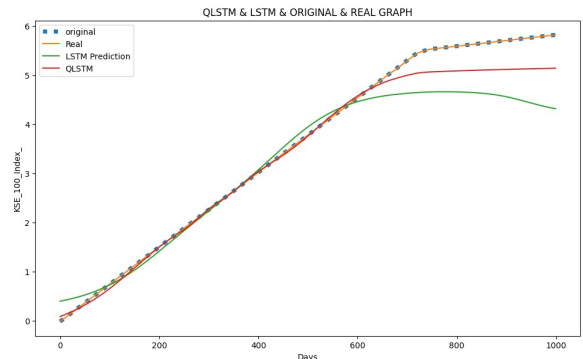

Figure 8: Comparison of the prediction of LSTM and QLSTM for the 1000 days real closed values of the KSE 100 index.

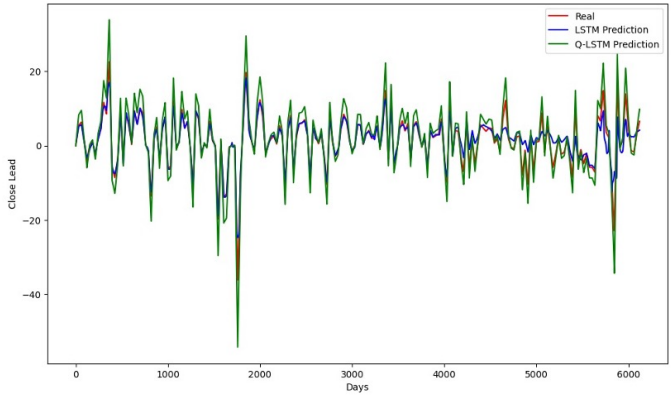

Figure 9: Comparison of the prediction of LSTM and QLSTM for real close lead values of the KSE 100 index.

Quantum Long Short-Term Memory (QLSTM) and Long Short-Term Memory (LSTM) have 26 input features and one output column. These models serve as powerful tools to establish correlations and capture intricate interdependencies and temporal patterns between the input and prediction columns. They are designed to learn from historical data, recognize patterns, and comprehend how the various input columns influence the output column. Both models undergo a training phase, during which they learn the underlying patterns by adjusting their internal parameters. This involves feeding the historical data into the models, comparing their predictions against the actual output, and optimizing their parameters to minimize the prediction error.

Both QLSTM and LSTM models thoroughly examine the 26 input columns to establish correlations with the prediction column, which signifies the exchange rate. Both models evaluate the interaction between input features and the output in a dynamic context.

These models hold immense potential in various fields, ranging from finance to healthcare and beyond. Uncovering correlations among the input features and prediction can assist in decision-making, risk assessment, trend forecasting, and more. The quantum element of QLSTM provides an additional dimension of computational power, potentially revealing insights that might have remained hidden in traditional models.

In essence, QLSTM and LSTM models, with their 26 input features and one output column, are valuable tools for exploring and establishing correlations within complex datasets. Whether leveraging the quantum power of QLSTM or the sophisticated temporal understanding of LSTM, these models offer a robust approach to unraveling relationships and making informed predictions.

Incorporating QLSTM and LSTM models into stock market prediction tasks offers a robust approach to analyzing complex financial data. While QLSTM exploits quantum advantages, LSTM excels in capturing sequential dependencies. The choice between these models hinges on the dataset's complexity and the prediction task's requirements. By leveraging their respective strengths, financial analysts and investors can potentially gain deeper insights into stock market behavior and make more informed decisions.

## 4. Conclusion

The present study identifies twenty-six economic, social, political, and administrative variables as predictors to predict the value of the KSE 100 index. Monthly data from February 2004 to December 2020 was taken for analysis. The mean absolute error loss of the LSTM and QLSTM during data training showed that the difference between actual and predicted values is minimal. Moreover, QLSTM has predicted values more accurately as compared to the LSTM. These results validate that the LSTM and QLSTM used in this study predict the stock market performance of a country based on not only macroeconomic or political indicators but also the inclusion of social and administrative indicators.

**Conflict of Interest:** The authors declare that there is no conflict of interest.
**Data Availability:** Data used in this study will be provided on request.
**Funding Statement:** The authors declare that they do not receive any funding from government or private organizations.
**Compliance with Ethical Standards:** This article does not contain any studies with human participants or animals performed by any of the authors.